\documentclass[review]{elsarticle}

\usepackage{hyperref}

\journal{Journal of \LaTeX\ Templates}

\begin{document}

\begin{frontmatter}

\title{On the event rate and luminosity function of superluminous supernovae}

\author[mymainaddress,mysecondaddress]{Wen-Chang Zhao}
\author[mymainaddress]{Xiao-Xin Xue}
\author[mythirdaddress]{Xiao-Feng Cao}
\cortext[mycorrespondingauthor]
{Corresponding author}
\ead{caoxf@mails.ccnu.edu.cn}

\address[mymainaddress]{Institute of Astrophysics, Central China Normal University,Wuhan, 430079,China}
\address[mysecondaddress]{Institute for Astronomy , Dali University, Dali, 671003, China}
\address[mythirdaddress]{School of Physics and Electronic Information, Hubei University of Education, Wuhan, 430205, China}

\begin{abstract}

We calculate the  rate per unit volume of hydrogen-poor superluminous supernovae (SLSNe-I) based on the 17 events discovered with the Pan-STARRS1 Medium Deep Survey (PS1 MDS). Two forms of the luminosity function (LF) are assumed : a log-normal form and a single power-law form, respectively. The rate of SLSNe-I is assumed to be proportional to the cosmic star formation rate with an additional redshift evolution of $(1+z)^{\alpha}$. Our results show that the single power-low form fits the data better than the log-normal form, and the event rate of SLSNe-I is proportional to the cosmic star formation rate directly (with $\alpha=0$). We measure the SLSNe-I  rate to be about $40 ~\rm{yr^{-1}Gpc^{-3} }$ at a weighted mean  redshift of  $\overline z=0.89$, which is consistent with  previous works.
\end{abstract}


\end{frontmatter}


\section{Introduction}

In the last decade, a class of superluminous transients was detected, which appear to be 10 - 100 times brighter than the normal core-collapse supernovae at  peak luminosity. They were labeled as Superluminous Supernovae (SLSNe) \citep{2012Sci...337..927G,2013ApJ...770..128I}, whose total radiated energy could be $10^{51}~\rm{erg}$. According to strength of the hydrogen emission lines in the spectrum, SLSNe can be  divided into two main types: hydrogen-rich SLSNe (SLSNe-II) and hydrogen-poor SLSNe (SLSNe-I). Most SLSNe-II exhibit narrow Balmer lines in their spectra, which are similar to these of Type IIn SNe. They are thought to be powered by the interaction of the blast wave with the dense circumstellar medium (CSM)  \cite{2007ApJ...659L..13O, 2011ApJ...735..106D, 2011ApJ...729..143C,2011ApJ...729...88R}. SLSNe-I are the H-poor SLSNe. They must be explosions of stellar cores stripped from their H-rich envelopes. Different from the SLSNe-\uppercase\expandafter{\romannumeral2}, the luminous emission of SLSNe-I is widely suggested to be powered by a newborn neutron star\cite{2010ApJ...717..245K, 2017MNRAS.470..197Y, 2017ApJ...840...12Y}.

Until 2018, there were about 30 SLSNe-I with both multiband photometric
light curves and complete spectra available in literatures \citep{2014ApJ...796...87I, 2015MNRAS.452.3869N}. Although the specimen number is relatively small, they still have been used to estimate the SLSNe-I event rate and explore the intrinsic luminosity function. For example, \cite{2014ApJ...796...87I}, where 16 SLSNe-I were used, suggested that a Gaussian form of the LF could fit the absolute peak magnitudes of SLSNe-I well within 1 $\sigma$ errors. \cite{2015MNRAS.452.3869N} also found that the magnitude distribution of SLSNe-I is  a Gaussian-like form with a low standard deviation. However, they argued that the Gaussian form can not necessarily represent their intrinsic LF because of the still small SLSNe-I number.
  It remains unclear whether SLSNe-I belong to the normal supernovae population or could be an independent population. The latest research shows that also low luminosity SLSNe-I may exist. It will be a challenge for a Gaussian-like form of the LF to explain these low luminosity SLSNe-I.
The  rates of SLSNe-I at different redshifts have been estimated by several authors
 \cite{2012Natur.491..228C,2013MNRAS.431..912Q,2015MNRAS.448.1206M,2017MNRAS.464.3568P,2015Natur.523..189G, 2019A&A...624A.143K}.

The uncertainties of the LF and the SLSNe-I  rate largely result from the fact that the analyzed data were combined with different detectors (whose observation field, response time and the limiting magnitude are diverse). Fortunately, \cite{2018ApJ...852...81L} present a new SLSNe-I sample from the Pan-STARRS1 Medium Deep Survey (PS1 MDS), which is an unified survey with consistent cadence and clear
footprint. Therefore, we will use this full and uniform data set to explore the event rate and the LF of SLSNe-I. We assume the SLSNe-I rate   trace the cosmic star formation history with an additional redshift evolution, e.g. $\propto (1+z)^\alpha$,
and two forms of LF are assumed: a log-normal one, and a single power-law one. By fitting the redshift and the luminosity distributions of SLSNe-I simultaneously with our model, we could constrain its event rate  and LF.

The paper is organized as follows. In section 2, we give an introduction to PS1 MDS and present the observational data. The luminosity threshold of PS1 MDS are also calculated. In section 3, models which apply to constrain the LF and the SLSNe-I rate are presented. Conclusions and discussions are given in Section 4.

\section{Observation of Pan-STARRS1 medium deep survey}

\subsection{Observational distribution of  SLSNe-I}

 We only use the 17 SLSNe-I from the  one single homogeneous Pan-STARRS1 Medium Deep Surveyz (PS1 MDS) to explore the SLSNe-I rate and the LF. These data   is taken from \cite{2018ApJ...852...81L}, where the criteria for screening the SLSNe and the redshifts of the SLSNes (via statistics and analysis of the four year PS1 MDS) are presented. The peak luminosity could be calculated by assuming  bolometric light curves for SLSNe-I, where a K-correction with the SNAKE method should be considered \citep{2018ApJ...852...81L, 2018MNRAS.475.1046I}. Following \cite{2018ApJ...852...81L}, we list their final results and the corresponding characteristics of the 17 SLSNe-I in Table 1, and the $z-L$ distribution of these 17 SLSNe-I is presented in Fig.1.

\subsection{ Detection threshold of PS1 telescope }

The PS1 is a wide field detection instrument, located at Haleakala, which mainly includes a 1.8 m diameter primary mirror and a series of pixel detectors \citep{Kaiser2010title,Tonry2009}. The 1.8 m PS1 telescope with the most advanced digital camera to date has a field of view of 7 square degrees. The PS1 MDS consists of ten fields (PS1 footprints) whose total field of view could reach 70 square degrees at high galactic latitudes spaced around the sky \citep{2016arXiv161205560C}. The PS1 MDS are observed in a multi-band with $gp1rp1ip1zp1yp1$ 5 bands. The depth of $gp1rp1ip1zp1$ in the PS1 telescope is about 23.3 mag (5$\sigma$), with $yp1$ a typical depth of $\sim21.7$ mag. For simplicity, we choose a typical limiting magnitude of $m=23$ for the whole bands to explore the observational threshold of the telescope.

The limiting magnitude of the PS1 can be regarded as the apparent magnitude of the faintest source that can be detected by the PS1 telescope. The relation between the absolute magnitude and the apparent magnitude can be written as:
\begin{equation}
M(z)=m+5-5\lg{d_L(z)},
\end{equation}
where $d_{L}$ is the luminosity distance of the source,
\begin{equation}
 d_{L}(z)=(1+z)c\int_{0}^{z}H(z')^{-1}dz',
\end{equation}
and the Hubble parameter
\begin{equation}
H(z)=H_{0} \sqrt{(1+z)^{3}\Omega_{m,0}+\Omega_{\Lambda,0}} ,
\end{equation}
where a flat $\Lambda$CDM universe is assumed, with the cosmology parameters:
 $\Omega_{m,0}=0.27 $ , $\Omega_{\Lambda,0}=0.73 $ , and
$H_{0}=70~\rm{km~s^{-1}~Mpc^{-1}}$.

The relation between the absolute magnitude and the luminosity can be written as:
\begin{equation}
 \lg{L(z)\over L_\odot}={1\over2.5}(M_\odot-M(z)),
\end{equation}
where $L_\odot$ and $M_\odot$ represent the luminosity and the
absolute magnitude of the Sun, respectively.

By substituting Eq.(1) into Eq.(4), we could drive the
observational luminosity threshold for PS1 by
\begin{equation}
 L_{\rm th}(z)=10^{{1\over2.5}(M_\odot-M(z))}L_\odot,
\end{equation}
which is shown as the solid red line in Fig.1.

\begin{figure}[ht!]
\centering
\includegraphics[scale=0.7]{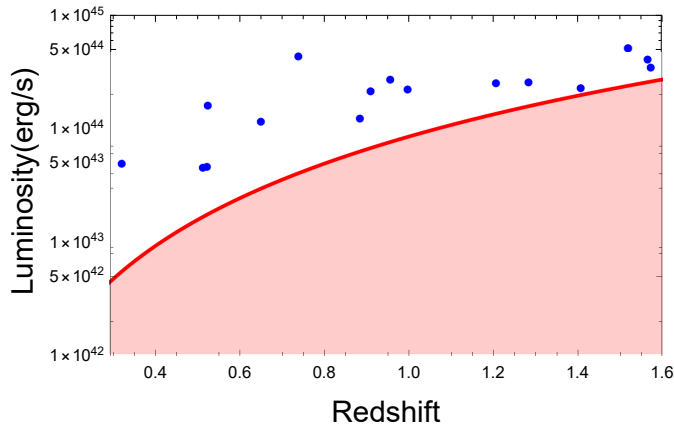}
\caption{Distribution of 17 SLSNe-I in a
$z-L$ diagram. The  solid line represents the observational luminosity
threshold of PS1 Telescope with the corresponding limit magnitude of
$m=23$.}
\end{figure}

\begin{table}[!ht]

\tabcolsep 4mm \caption{ 17 SLSNe-I with measure redshift}
\begin{center}
\begin{tabular}
{cccccccc}
\hline \multicolumn{2}{c}{Object}
&\multicolumn{2}{c}{Redshift}
&\multicolumn{2}{c}{$T_{BB}$ at peak } &\multicolumn{2}{c}{Peak Lum }\\
   &\   &\   &\   &[$K$]&    &[$10^{44}~\rm{erg ~ s^{-1}}$]

\\ \hline
&PS1-12cil     &0.32&   &13000&   &0.50      \\
&PS1-14bj    &0.5125&   &7000&   &0.46 \\
&PS1-12bqf    &0.522&   &11000&   &0.47       \\
& PS1-11ap   &0.524&  &10000&   &1.63   \\
&  PS1-10bzj   &0.65&   &17000&  &1.17      \\
& PS1-11bdn   &0.738&  &16000& &4.39  \\
& PS1-13gt   &0.884&    &6000&  &1.25     \\
&PS1-10awh   &0.909& &16000&  &2.16      \\
& PS1-10ky   &0.956&   &16000&   &2.75  \\
&PS1-11aib    &0.997&   &10000&   &2.24       \\
& PS1-10ahf   &1.10 &  &10000&   &1.137     \\
&  PS1-10pm   &1.206&   &8000&   &2.56     \\
& PS1-11tt   &1.283&   &9000&   &2.61 \\
&PS1-11afv    &1.407&  &12000&  &2.32   \\
&PS1-13or    &1.52&  &11000&    &5.20  \\
&PS1-11bam    &1.565&  &12000&  &4.13   \\
&PS1-12bmy    &1.572&  &9000&    &3.51  \\
\hline
\end{tabular}
\footnotetext[1]{Calculated with $\log ft=5.3$, $T_{9}=0.45$ and
$\dot{m}=0.1$.} \footnotetext[2]{Calculated with $\log ft=5.3$,
$T_{9}=0.45$ and $\dot{m}=0.2$.}
\end{center}
\end{table}

\section{The model and results}

\subsection{Luminosity function of SLSNe-I}

In the previous works, the LF of SLSNe-I is usually assumed to be a Gaussian-like form, which is the same as that for the normal supernovae \citep{2014ApJ...796...87I}. Observations from the subsequent studies also support this result ( e.g., \cite{2015MNRAS.452.3869N, 2018ApJ...852...81L}). However, recent studies suggest that low luminosity SLSNe-I may  exist \citep{2018ApJ...852...81L,2018ApJ...860..100D}, which will make the form of the LF to be different. Therefore, the actual form of the LF may be far from a Gaussian distribution. Following the studies of the LF of long gamma-ray burst (LGRB) and the intrinsic energy function of fast radio bursts (FRB) (e.g., \cite{2013ApJ...772L...8T} for GRBs, \cite{2017RAA....17...14C} for FRBs), we
suggest that a single power-law form LF for SLSNe-I is possible and empirically valid. Based on the current observations and theoretical analysis, we introduce a log-normal form and a power-law form LF for SLSNe-I. By fitting the distributions of the observed data with our
model, we try to explore the intrinsic LF. The specific expressions of the two LF are as follows:

1) The log-normal form of the LF for SLSNe-I:

\begin{equation}
\Phi_{1}(L)\propto \exp[{\frac{-(\lg{L_{p}}-\lg
{L_{\mu}})^{2}}{2\sigma^{2}}}],
\end{equation}
where $L_{\mu}$ and $\sigma$ are the free parameters determined by the intrinsic luminosity distribution of SLSNe-I, $L_{\mu}$ represents the average peak luminosity, $\sigma$ represents the standard deviation of the log-normal distribution.

2) The single-power-law form:
\begin{equation}
\Phi_{2}(L)\propto  L^{-\gamma},
\end{equation}
where the index $\gamma$ is a free parameter.

\subsection{The  rate of SLSNe-I }

 As supernova explosions mark the  death of a star,  SLSNe are expected to be a tracer of the star formation, and  we therefore expect the SLSNe-I burst rate to trace the star formation rate. To be specific, we assume the SLSNe-I burst rate $\dot{R}(z)$  to be proportional to the star formation rate $\dot{\rho_{*}}(z)$, where an additional redshift evolution should be considered:
\begin{equation}
  \dot{R}(z)=C(1+z)^{\alpha} \dot{\rho_{*}}(z),
\end{equation}
here $\alpha$ is a free parameter, $C$ is a proportionality constant.

The additional redshift evolution is determined by  complicated physical conditions. For example, the special requirements for the origin of SLSNe-I, e.g., mass, metalicity and
magnetic field of its progenitor (the same as that of the long GRBs; \cite{2006ApJ...638L..63L,2011MNRAS.416.2174C}), the evolution of the LF, the effects of the time delay, and the observation efficiency of the telescope. The star formation rates at relatively low redshifts, which have been accurately measured \citep{2006ApJ...651..142H}, can be described by
\begin{equation}
   \dot{\rho_{*}}(z)\propto
   \left \{
   \begin{array}{l}
  (1+z)^{3.44},     \quad z<0.97, \\
   (1+z)^{-0.26},  \quad 0.97\le z<4,
   \end{array}
   \right.
\end{equation}
with the local star formation rate $ \dot{\rho_{*}}(0)=0.02 M_\odot~\rm{yr^{-1}~Gpc^{-3}}$.

\subsection{Results}

The theoretical cumulative SLSNe-I number within a special redshift range is a function of the LF and the  rate of SLSNe-I. Combing  Eqs.(5)-(9), the number of SLSNe-I within the redshift range, $z_{\rm min}<z<z_{\rm max}$ can be described by
\begin{equation}
N^{\rm{obs}}_{<z}= \mathcal T \frac{\mathcal A}{4\pi} \int^{z_{\rm max}}_{z_{\rm min}}\int^{L_{\rm max}}_
{L_{\rm th}(z)}\Phi_{i}(L)\dot{R}(z)dL\frac{dV(z^{'})}{1+z^{'}},
\end{equation}
where $\Phi_{i}(L)$ represents different forms of the LF, $
{\mathcal A}/{4\pi}\sim1.7\times10^{-3}$ is the ratio between the total field view
of PS1 MDS to the whole sky area, $\mathcal T\sim4~\rm yr$ is the
total observation time period, $dV(z)$ is the comoving volume
element, and $dV(z)/dz=4\pi d_{c}(z)^{2}c/H(z)$, $(1+z)$ represents
the time dilation of the universe. In our calculation, we set
$L_{\rm max}=10^{45}$ erg s$^{-1}$. The cumulative number within a
luminosity range of $L_{\rm min}<L<L_{\rm max}$ is describe by
\begin{equation}
N^{\rm obs}_{<L}= \mathcal T \frac{\mathcal A}{4\pi} \int^{L_{\rm max}}_{L_{\rm min}}\int^{z_{\rm{max}}}_{z_{\rm{min}}}\Phi_{i}(L)\dot{R}(z)\frac{dV(z^{'})}{1+z^{'}}dL,
\end{equation}
where we set the minimum luminosity at  $L(0.5)$.

Combing Eqs.(10) and(11), we could derive the expected redshift
and luminosity distributions of SLSNe-I. By iteratively changing the
free parameters mentioned above, some plausible fittings can be
found. For the log-normal LF, we have tried all the possible
parameters in a permissible parameter range, but no valid parameters
 could be found. We show a feasible fitting with a log-normal LF in
Fig.2 (with the specific parameter values of $\lg L_{\mu}=44.25$,
$\sigma=0.3344$ ). For the single power-law form of the LF, we find
that the free parameter of $\gamma$ in the range of -2 to 1 is
available, and the parameter $\alpha$ within -2.5 and 2 is
acceptable. To get the best fitted values of $\alpha$ and $\gamma$,
we further fitted the cumulative number distributions of redshift
and luminosity, simultaneously. Some feasible fittings with special
parameter values are shown in Figs. 3, 4 and 5.

We find that the value of $\gamma$ in the single power-law form LF
has greater impacts on the distributions than the redshift evolution
of $\alpha$. Therefore, we set two values of $\alpha$ ($\alpha=0, -2.5$
) and several more specific values of $\gamma$
($\gamma=0.5$, $-1$, $-1.7$), respectively, to find the best fitting
results.

\begin{figure}
\centering\resizebox{\hsize}{!}{\includegraphics{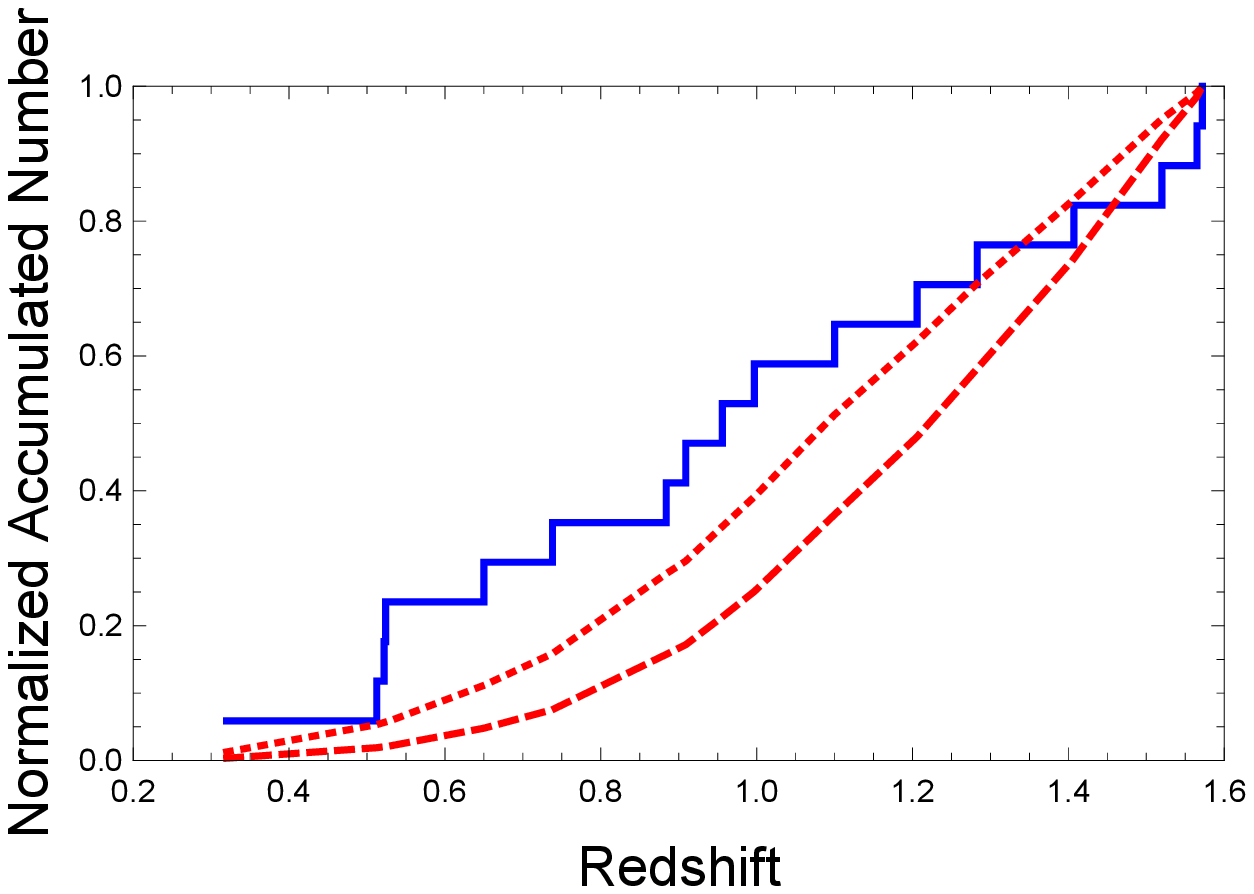},\includegraphics{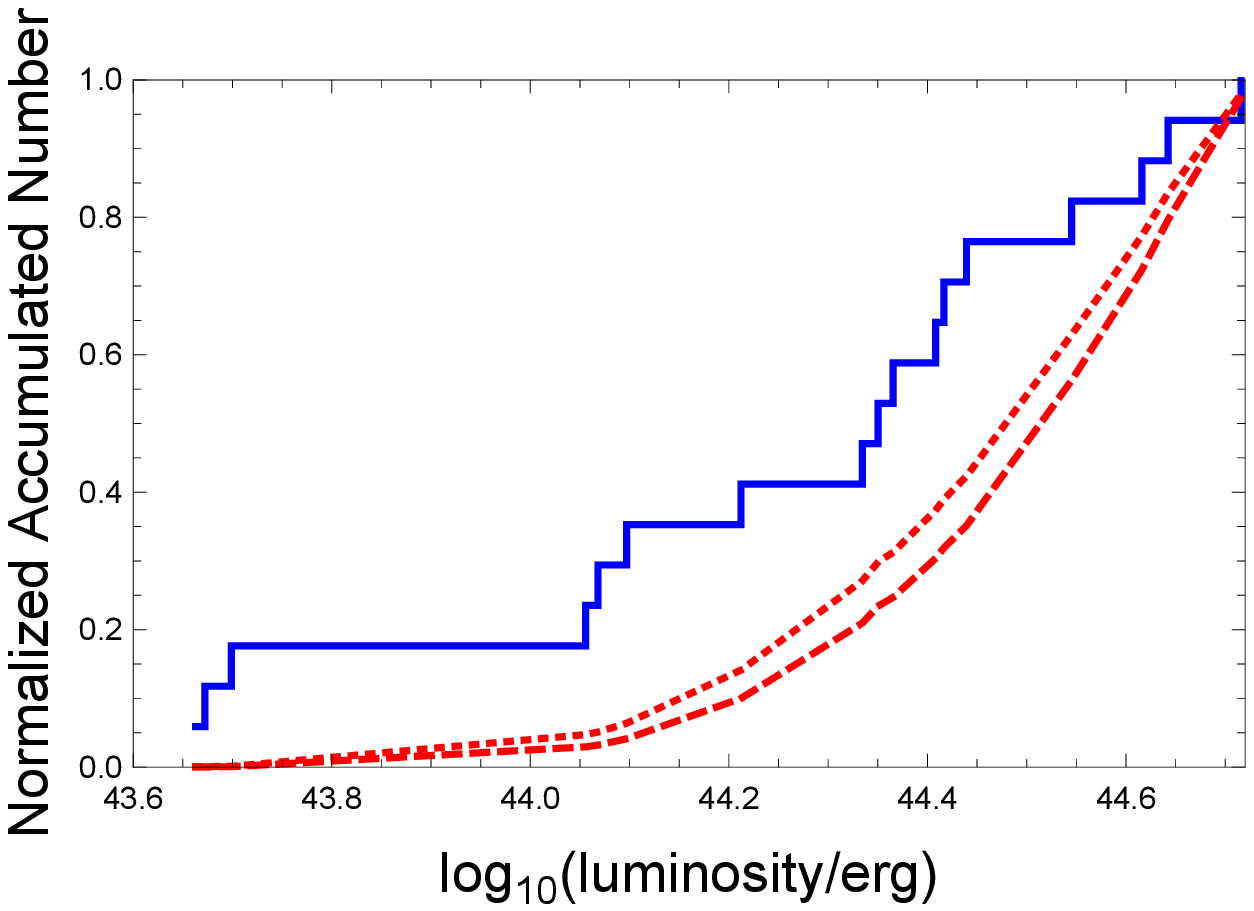}}
\caption{Normalized cumulative redshift distribution and
luminosity distribution of SLSNe-I for $\alpha=-2.5$ (dashed line) and
$\alpha=0$ (dotted line), respectively. The LF is assumed to be a
log-normal form and the corresponding values of the parameters are
$\lg L_{\mu}=44.25$, $\sigma=0.3344$. }
   \label{fig:2}
\end{figure}

\begin{figure}
\centering\resizebox{\hsize}{!}{\includegraphics{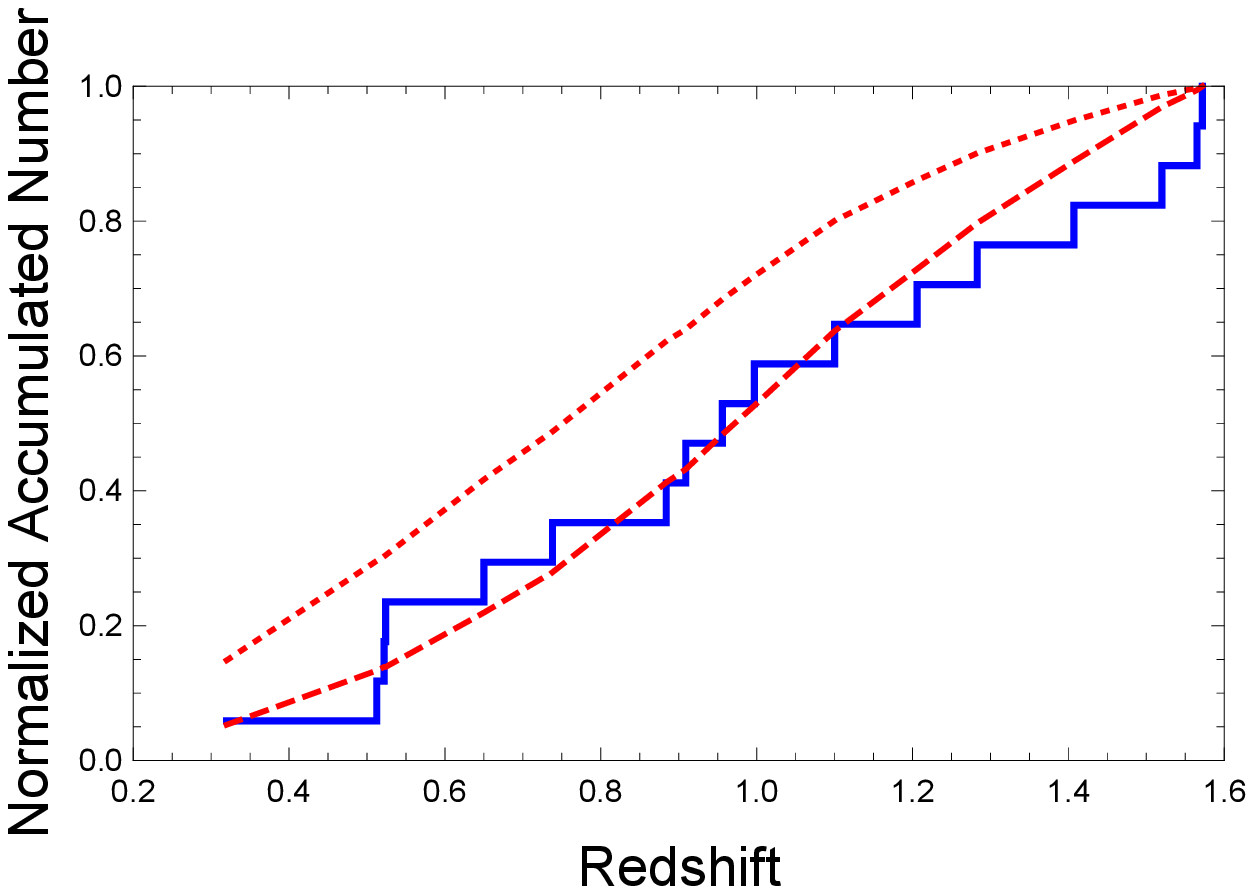},\includegraphics{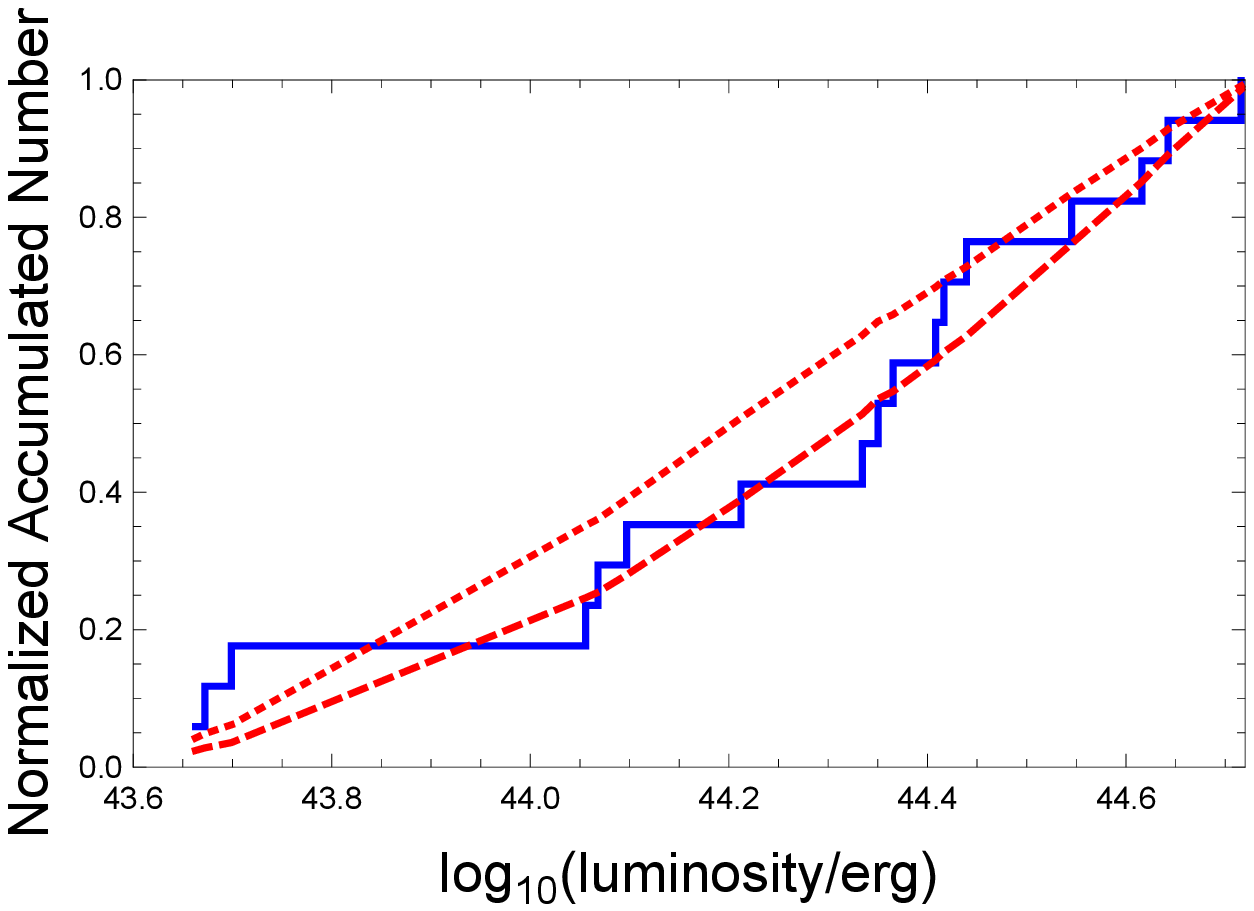}}
\caption{The normalized cumulative redshift distribution and
luminosity distribution for the single power-law LF with
$\gamma=-1.7$,  the dotted and dashed lines are for $\alpha=-2.5$ and
$\alpha=0$, respectively. }
   \label{fig:3}
\end{figure}

\begin{figure}
\centering\resizebox{\hsize}{!}{\includegraphics{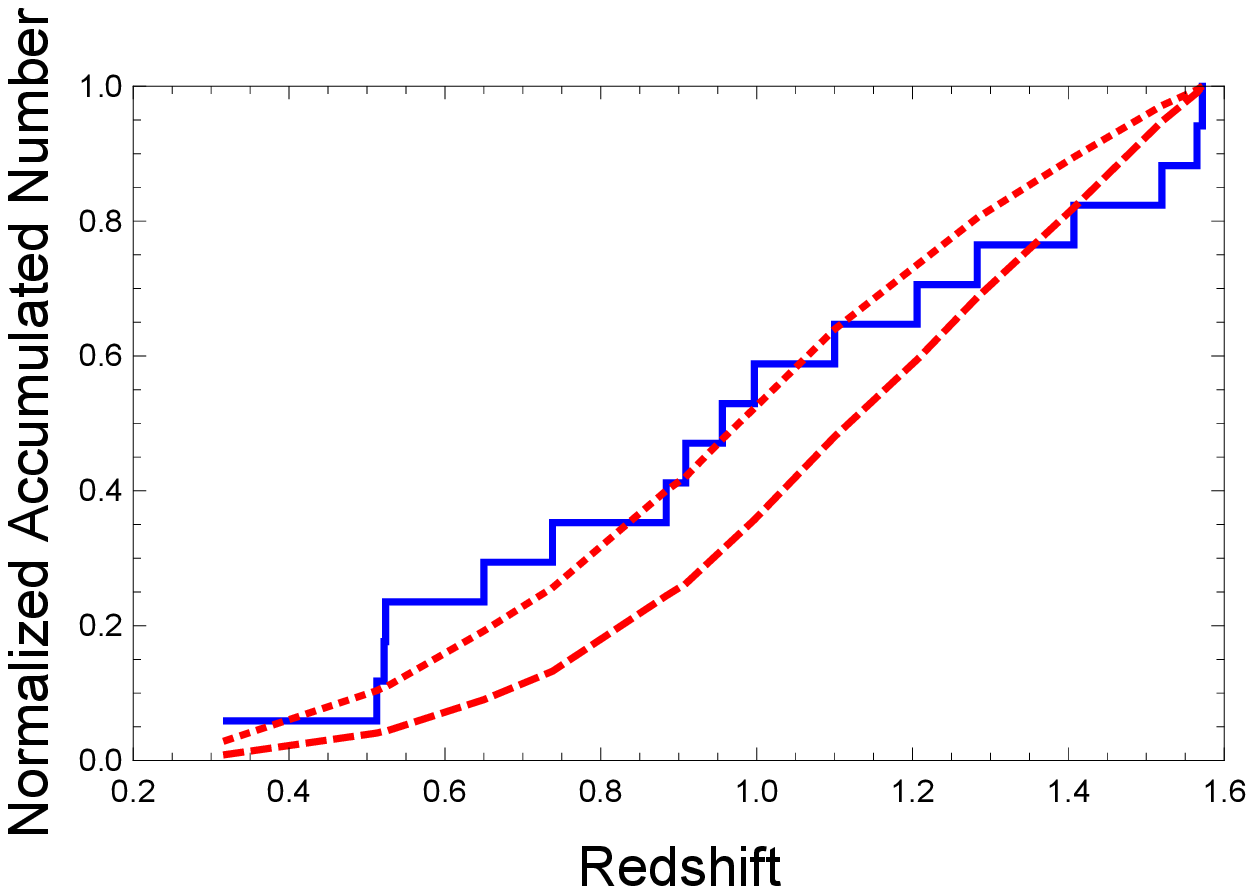},\includegraphics{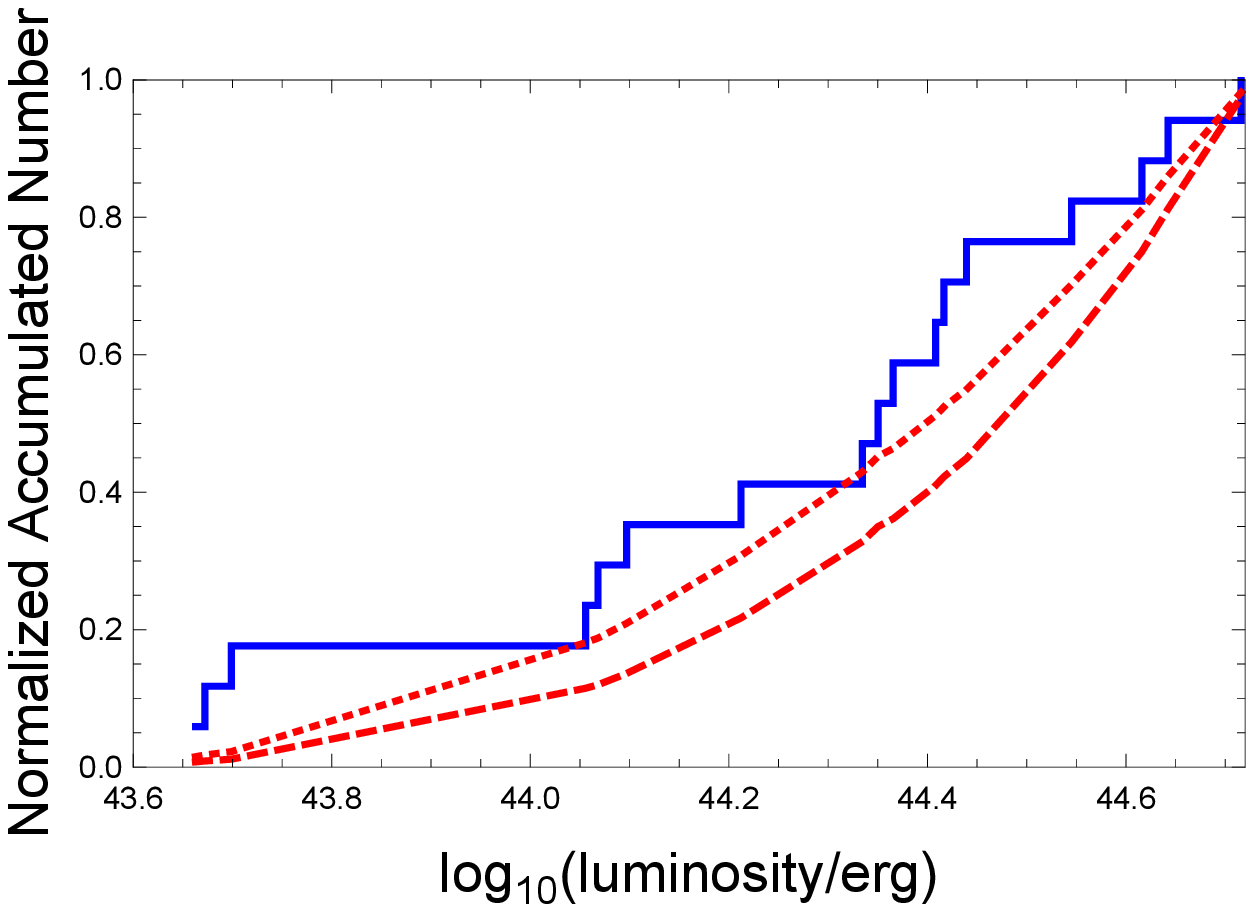}}
\caption{The same as Fig.3, with $\gamma=-1$ } \label{fig:4}
\end{figure}

\begin{figure}
\centering\resizebox{\hsize}{!}{\includegraphics{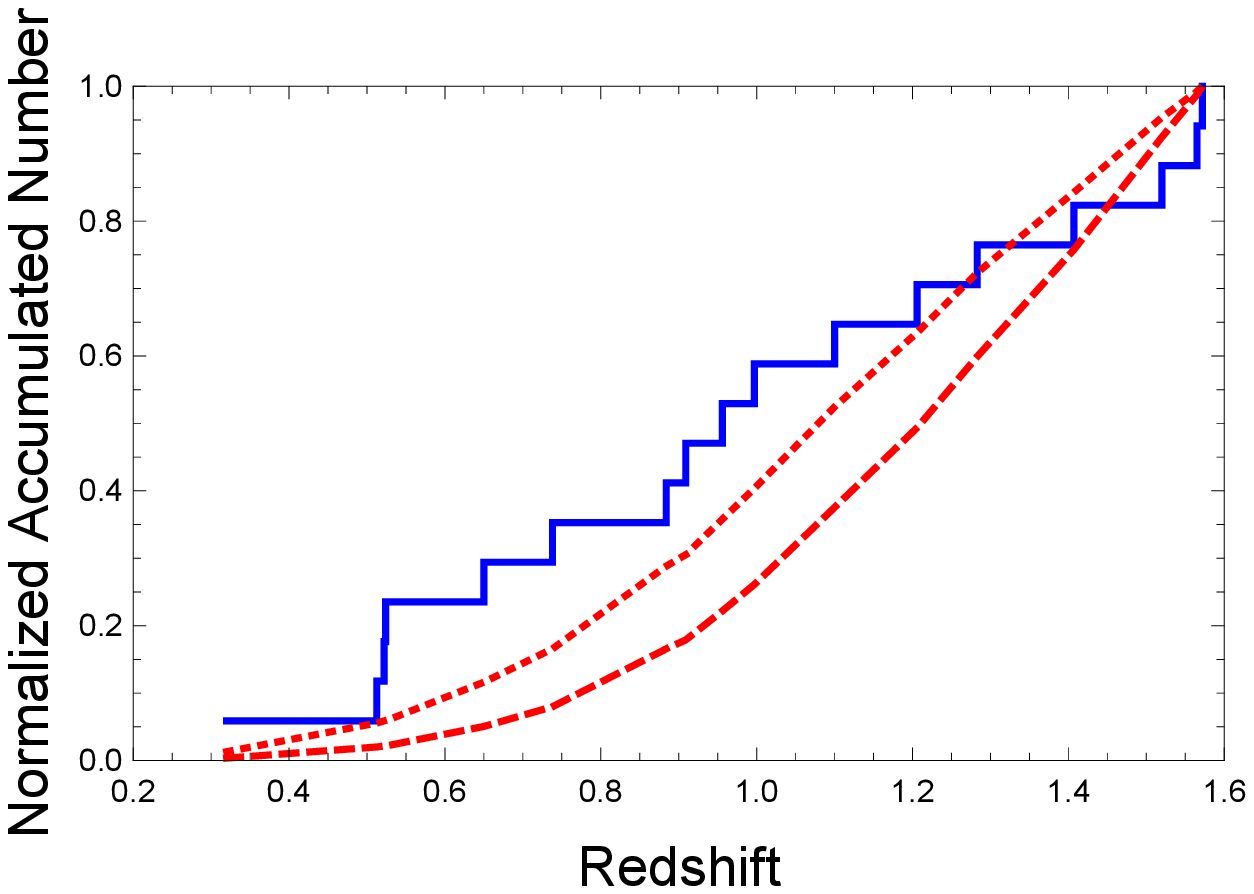},\includegraphics{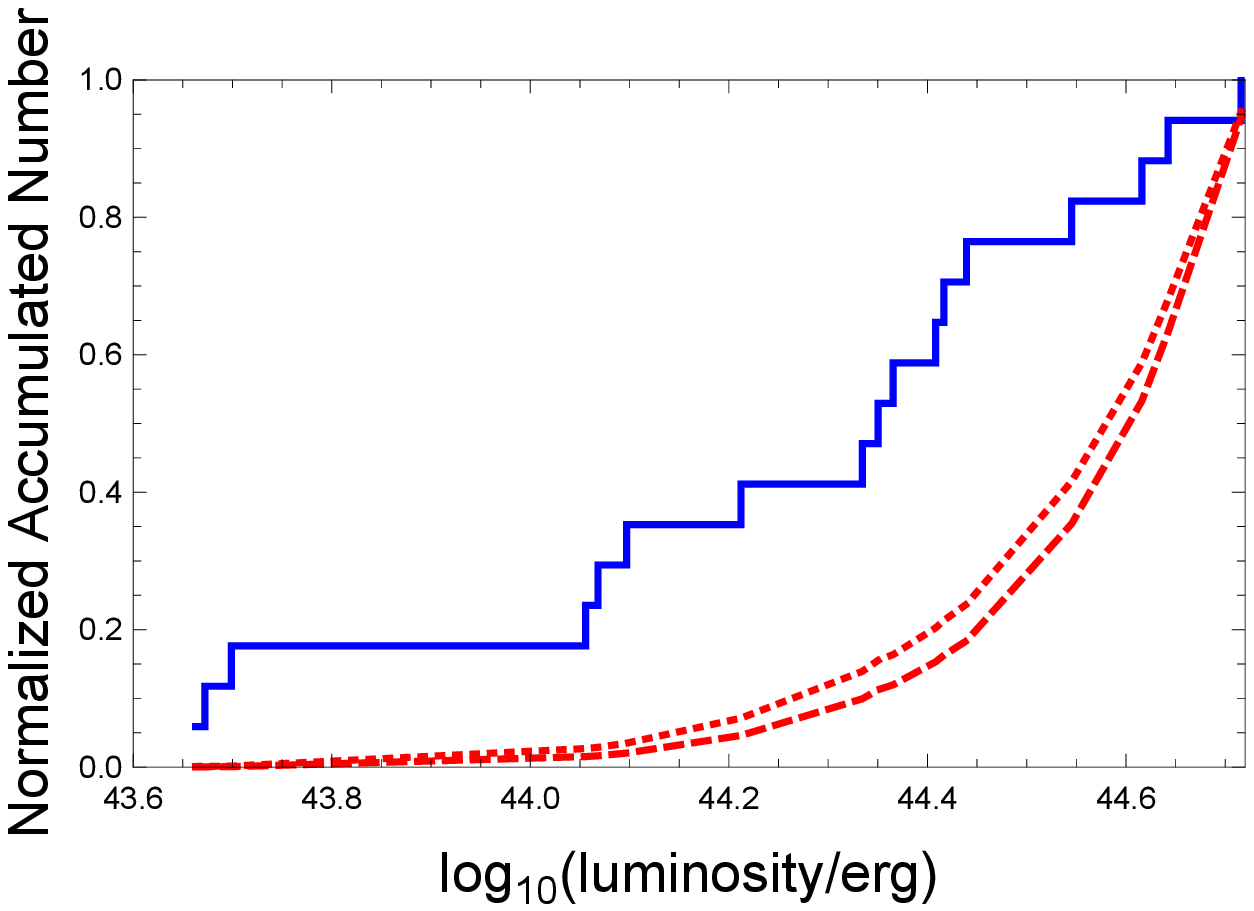}}
\caption{The same as Fig.4, with $\gamma=0.5$  }
   \label{fig:5}
\end{figure}

\begin{table*}
 \centering
 \caption{Parameters and test results}
 \label{symbols}
 \resizebox{\textwidth}{15mm}{
 \begin{tabular}{c|c|c|c|c|c|c|c|c}
  \hline
	Parameter& $\alpha=-2.5$ & $\alpha=0$& $\alpha=-2.5$& $\alpha=0$& $\alpha=-2.5$& $\alpha=0$& $\alpha=-2.5$& $\alpha=0$\\
&$\gamma=-1.7$ &$\gamma=-1.7$&$\gamma=-1$&$\gamma=-1$&$\gamma=0.5$&$\gamma=0.5$&$\rm log-normal$ &$\rm log-normal$             \\
 \hline
   $P_{\rm test}(z)$ & 0.90986   &  0.999805   &  0.995959  & 0.722396  &   0.832157  &   0.576525  &  0.74805  &  0.565005  \\
\hline

$P_{\rm test}(L)$ & 0.989138   &  0.980619   &  0.846204  & 0.617766  &   0.503399  &   0.50118  &  0.560665  &   0.518076  \\
 \hline
 \end{tabular}}
 \end{table*}

We use the AndersonDarlingTest to get the best fitting results. The
values of the fitting parameters and the goodness of the fits (both
in redshift and luminosity) are shown in Table $2$, where a higher
$P$ value represents a better fitting result.

The best fitting results for both redshift and luminosity
distributions are obtained (with $P_{\rm test}(z)=0.999805$ and
$P_{\rm test}(L)=0.989138$ ), and the best fitting values of the
parameters are $\gamma=-1.7$ and $\alpha=0$, as shown in Table $2$.
It is obvious that the single power-law LF fits the data better than
the log-normal LF, and it seems that the SLSNe-I burst rate follows
the cosmic star formation rate directly without a redshift evolution
(with $\alpha=0$).

With the best fitting parameters of $\gamma=-1.7$ and $\alpha=0$
given above, we normalized Eq.(7) with $\gamma=-1.7$ and equal
Eq.(10) (with $z_1=1.6$ and $\alpha=0$) to the observed number
of SLSNe-I ($N_{\rm obs}=17$), the constant coefficient $C$ between the
SLSNe-I rate and the star formation rate can be well determined.
Finally, the SLSNe-I rate is estimated to be around
$40~\rm{yr^{-1}~Gpc^{-3}}$ at redshift $\overline z=0.89$ (the volume-weighted centre of the $0<z<1.6 $ range), which is
consistent with the results of previous works.

\section{Summary and discussion }

 We assumed that SLSNe-I rate is proportional to the star
formation rate with an additional redshift evolution of
$(1+z)^\alpha$ and the LF takes the form of log-normal form and a
single-power-law, respectively. By fitting the distribution of the 17 SLSNe-I detected by Pan-STARRS1
both of redshift and luminosity with our model, a rough valid
parameter space of single-power-law form LF is obtained. On the other hand the log-normal one does not fit. This result indicates
that the intrinsic LF of SLSNe-I is more likely to be a
single-power-law form rather than log-normal form, which
could be supported by the future observations. Furthermore, we
were able to narrow down the parameters in the single-power-law form LF and the SLSNe-I
rate, the result being the best fitting (with parameters $\gamma=-1.7$ and $\alpha=0$) which illustrates that the  rate of SLSNe-I is proportional to the
cosmic star formation rate directly. This result may highlight that
the  SLSNe-I  explosions originate from the death of some peculiar
type of massive stars. Based on the parameters of the best fit, we
further estimate the event rate of SLSNe-I as
$40~\rm{yr^{-1}~Gpc^{-3}}$ at a volume-weighted redshift of $\overline z=0.89$,
which is well
consistent with  the current literature. In future analysis, with a more
accurate SLSNe-I rate and a magnetar formation rate, the ratio of the
magnetic star in the SLSNe-I could be estimated, which can be used to
independently constrain the magnetar model of SLSNe-I.

\section*{Acknowledgements}

The authors thank  Yun-Wei Yu and Wei-Wei Tan very much for helpful comments and suggestions which have significant support to our work. We are also very grateful for the valuable suggestions and comments of the referee, which improved the manuscript significantly. This work is supported by the National Natural Science Foundation of China (grant Nos. U1838203 and 11863002) and by the opening project of the Key Laboratory of Quark and Lepton Physics (MOE) at CCNU (grant no. QLPL 2018P01)

\section*{References}

\bibliography{slsnv2}

\begin{thebibliography}{10}
\expandafter\ifx\csname url\endcsname\relax
  \def\url#1{\texttt{#1}}\fi
\expandafter\ifx\csname urlprefix\endcsname\relax\def\urlprefix{URL }\fi
\expandafter\ifx\csname href\endcsname\relax
  \def\href#1#2{#2} \def\path#1{#1}\fi

\bibitem{2012Sci...337..927G}
A.~{Gal-Yam}, {Luminous supernovae}, Science 337~(6097) (2012) 927.
\newblock \href {http://dx.doi.org/10.1126/science.1203601}
  {\path{doi:10.1126/science.1203601}}.

\bibitem{2013ApJ...770..128I}
C.~{Inserra}, S.~J. {Smartt}, A.~{Jerkstrand}, S.~{Valenti}, M.~{Fraser},
  et~al., {Super-luminous Type Ic supernovae: catching a magnetar by the tail},
  apj 770~(2) (2013) 128.

\bibitem{2007ApJ...659L..13O}
E.~O. {Ofek}, P.~B. {Cameron}, M.~M. {Kasliwal}, A.~{Gal-Yam}, et~al., {SN
  2006gy: An extremely luminous supernova in the galaxy NGC 1260}, apjl 659~(1)
  (2007) L13--L16.
\newblock \href {http://dx.doi.org/10.1086/516749} {\path{doi:10.1086/516749}}.

\bibitem{2011ApJ...735..106D}
A.~J. {Drake}, S.~G. {Djorgovski}, A.~{Mahabal}, et~al., {The discovery and
  nature of the optical transient CSS100217:102913+404220}, apj 735~(2) (2011)
  106.
\newblock \href {http://dx.doi.org/10.1088/0004-637X/735/2/106}
  {\path{doi:10.1088/0004-637X/735/2/106}}.

\bibitem{2011ApJ...729..143C}
E.~{Chatzopoulos}, J.~C. {Wheeler}, J.~{Vinko}, et~al., {SN 2008am: a
  super-luminous Type IIn supernova}, apj 729~(2) (2011) 143.
\newblock \href {http://dx.doi.org/10.1088/0004-637X/729/2/143}
  {\path{doi:10.1088/0004-637X/729/2/143}}.

\bibitem{2011ApJ...729...88R}
A.~Rest, R.~J. Foley, S.~Gezari, G.~Narayan, et~al., Pushing the boundaries of
  conventional core-collapse supernovae: the extremely energetic supernova sn
  2003ma, apj 729~(2) (2011) 88.
\newblock \href {http://dx.doi.org/10.1088/0004-637X/729/2/88}
  {\path{doi:10.1088/0004-637X/729/2/88}}.

\bibitem{2010ApJ...717..245K}
D.~{Kasen}, L.~{Bildsten}, {Supernova light curves powered by young magnetars},
  apj 717~(1) (2010) 245.
\newblock \href {http://dx.doi.org/10.1088/0004-637X/717/1/245}
  {\path{doi:10.1088/0004-637X/717/1/245}}.

\bibitem{2017MNRAS.470..197Y}
Y.-W. {Yu}, S.-Z. {Li}, {A possible relation between flare activity in
  super-luminous supernovae and gamma-ray bursts}, mnras 470~(1) (2017) 197.
\newblock \href {http://dx.doi.org/10.1093/mnras/stx1028}
  {\path{doi:10.1093/mnras/stx1028}}.

\bibitem{2017ApJ...840...12Y}
Y.-W. {Yu}, J.-P. {Zhu}, S.-Z. {Li}, H.-J. {L{\"u}}, Y.-C. {Zou}, {A
  statistical study of superluminous supernovae using the magnetar engine model
  and implications for their connection with gamma-ray bursts and hypernovae},
  apj 840~(1) (2017) 12.
\newblock \href {http://dx.doi.org/10.3847/1538-4357/aa6c27}
  {\path{doi:10.3847/1538-4357/aa6c27}}.

\bibitem{2014ApJ...796...87I}
C.~{Inserra}, S.~J. {Smartt}, {Superluminous supernovae as standardizable
  candles and high-redshift distance probes}, apj 796~(2) (2014) 87.
\newblock \href {http://dx.doi.org/10.1088/0004-637X/796/2/87}
  {\path{doi:10.1088/0004-637X/796/2/87}}.

\bibitem{2015MNRAS.452.3869N}
M.~{Nicholl}, S.~J. {Smartt}, A.~{Jerkstrand}, C.~{Inserra}, et~al., {On the
  diversity of superluminous supernovae: ejected mass as the dominant factor},
  mnras 452~(4) (2015) 3869--3893.
\newblock \href {http://dx.doi.org/10.1093/mnras/stv1522}
  {\path{doi:10.1093/mnras/stv1522}}.

\bibitem{2012Natur.491..228C}
J.~{Cooke}, M.~{Sullivan}, A.~o. {Gal-Yam}, {Superluminous supernovae at
  redshifts of 2.05 and 3.90}, nat 491~(7423) (2012) 228--231.
\newblock \href {http://dx.doi.org/10.1038/nature11521}
  {\path{doi:10.1038/nature11521}}.

\bibitem{2013MNRAS.431..912Q}
R.~M. {Quimby}, F.~{Yuan}, C.~{Akerlof}, J.~C. {Wheeler}, {Rates of
  superluminous supernovae at z {\ensuremath{\sim}} 0.2}, mnras 431~(1) (2013)
  912--922.
\newblock \href {http://dx.doi.org/10.1093/mnras/stt213}
  {\path{doi:10.1093/mnras/stt213}}.

\bibitem{2015MNRAS.448.1206M}
M.~{McCrum}, S.~J. {Smartt}, A.~{Rest}, K.~{Smith}, et~al., {Selecting
  superluminous supernovae in faint galaxies from the first year of the
  Pan-STARRS1 Medium Deep Survey}, mnras 448~(2) (2015) 1206--1231.
\newblock \href {http://arxiv.org/abs/1402.1631} {\path{arXiv:1402.1631}},
  \href {http://dx.doi.org/10.1093/mnras/stv034}
  {\path{doi:10.1093/mnras/stv034}}.

\bibitem{2017MNRAS.464.3568P}
S.~{Prajs}, M.~{Sullivan}, M.~{Smith}, A.~{Levan}, et~al., {The volumetric rate
  of superluminous supernovae at z {\ensuremath{\sim}} 1}, mnras 464~(3) (2017)
  3568--3579.
\newblock \href {http://dx.doi.org/10.1093/mnras/stw1942}
  {\path{doi:10.1093/mnras/stw1942}}.

\bibitem{2015Natur.523..189G}
J.~{Greiner}, P.~A. {Mazzali}, D.~A. {Kann}, et~al., {A very luminous
  magnetar-powered supernova associated with an ultra-long
  {\ensuremath{\gamma}}-ray burst}, nat 523~(7559) (2015) 189--192.
\newblock \href {http://dx.doi.org/10.1038/nature14579}
  {\path{doi:10.1038/nature14579}}.

\bibitem{2019A&A...624A.143K}
D.~A. {Kann}, P.~{Schady}, F.~{Olivares E.}, et~al., {Highly luminous
  supernovae associated with gamma-ray bursts. I. GRB 111209A/SN 2011kl in the
  context of stripped-envelope and superluminous supernovae}, aap 624 (2019)
  A143.
\newblock \href {http://dx.doi.org/10.1051/0004-6361/201629162}
  {\path{doi:10.1051/0004-6361/201629162}}.

\bibitem{2018ApJ...852...81L}
R.~{Lunnan}, R.~{Chornock}, E.~{Berger}, D.~O. {Jones}, et~al., {Hydrogen-poor
  superluminous supernovae from the Pan-STARRS1 medium deep survey}, apj
  852~(2) (2018) 81.
\newblock \href {http://dx.doi.org/10.3847/1538-4357/aa9f1a}
  {\path{doi:10.3847/1538-4357/aa9f1a}}.

\bibitem{2018MNRAS.475.1046I}
C.~{Inserra}, S.~J. {Smartt}, E.~E.~E. {Gall}, et~al., {On the nature of
  hydrogen-rich superluminous supernovae}, mnras 475~(1) (2018) 1046--1072.
\newblock \href {http://dx.doi.org/10.1093/mnras/stx3179}
  {\path{doi:10.1093/mnras/stx3179}}.

\bibitem{Kaiser2010title}
N.~Kaiser, W.~Burgett, K.~Chambers, L.~Denneau, J.~Tonry, The pan-starrs
  wide-field optical/nir imaging survey, Proceedings of SPIE - The
  International Society for Optical Engineering 7733~(6) (2010) --.

\bibitem{Tonry2009}
J.~Tonry, P.~Onaka, The pan-starrs gigapixel camera, in: Advanced Maui Optical
  and Space Surveillance Technologies CSonference, 2009, p. E40.

\bibitem{2016arXiv161205560C}
K.~C. {Chambers}, E.~A. {Magnier}, N.~{Metcalfe}, et~al., {The Pan-STARRS1
  surveys}, arXiv e-prints (2016) arXiv:1612.05560.

\bibitem{2018ApJ...860..100D}
A.~{De Cia}, A.~{Gal-Yam}, A.~{Rubin}, et~al., {Light curves of hydrogen-poor
  superluminous supernovae from the palomar transient factory}, apj 860~(2)
  (2018) 100.
\newblock \href {http://dx.doi.org/10.3847/1538-4357/aab9b6}
  {\path{doi:10.3847/1538-4357/aab9b6}}.

\bibitem{2013ApJ...772L...8T}
W.-W. Tan, X.-F. Cao, Y.-W. Yu, Determining the luminosity function of swift
  long gamma-ray bursts with pseudo-redshifts, apjl 772~(1) (2013) L8.
\newblock \href {http://dx.doi.org/10.1088/2041-8205/772/1/L8}
  {\path{doi:10.1088/2041-8205/772/1/L8}}.

\bibitem{2017RAA....17...14C}
X.-F. {Cao}, M.~{Xiao}, F.~{Xiao}, {Modeling the redshift and energy
  distributions of fast radio bursts}, Res. in Astron. and Astrophys. 17~(2)
  (2017) 14.
\newblock \href {http://dx.doi.org/10.1088/1674-4527/17/2/14}
  {\path{doi:10.1088/1674-4527/17/2/14}}.

\bibitem{2006ApJ...638L..63L}
N.~{Langer}, C.~A. {Norman}, {On the collapsar model of long gamma-ray bursts:
  constraints from cosmic metallicity evolution}, apjl 638~(2) (2006) L63--L66.
\newblock \href {http://dx.doi.org/10.1086/500363} {\path{doi:10.1086/500363}}.

\bibitem{2011MNRAS.416.2174C}
X.-F. {Cao}, Y.-W. {Yu}, K.~S. {Cheng}, X.-P. {Zheng}, {The luminosity function
  of Swift long gamma-ray bursts}, mnras 416~(3) (2011) 2174--2181.
\newblock \href {http://dx.doi.org/10.1111/j.1365-2966.2011.19194.x}
  {\path{doi:10.1111/j.1365-2966.2011.19194.x}}.

\bibitem{2006ApJ...651..142H}
A.~M. {Hopkins}, J.~F. {Beacom}, {On the normalization of the cosmic star
  formation history}, apj 651~(1) (2006) 142--154.
\newblock \href {http://dx.doi.org/10.1086/506610} {\path{doi:10.1086/506610}}.

\end{thebibliography}

\end{document}